\documentclass[showpacs,aps,pra,twocolumn,10pt,superscriptaddress]{revtex4-1}%
\usepackage{amssymb}
\usepackage{graphicx}
\usepackage{dcolumn}
\usepackage{amsmath}
\usepackage{relsize}
\usepackage{subfigure}
\usepackage{CJK,color}
\usepackage{amsfonts}%
\providecommand{\U}[1]{\protect\rule{.1in}{.1in}}
\begin{document}
\begin{CJK*}{GBK}{song} 
\title{Strong mechanical squeezing in an unresolved-sideband optomechanical system}

\author{Rong Zhang}
\address{Institute of Theoretical Physics, Chinese Academy of Sciences, Beijing 100190, China}
\address{School of Physical Sciences, University of Chinese Academy of Sciences, No.19A Yuquan Road, Beijing 100049, China}

\author{Yinan Fang}
\address{Beijing Computational Science Research Center, Beijing 100193, China}

\author{Yang-Yang Wang}
\address{Institute of Theoretical Physics, Chinese Academy of Sciences, Beijing 100190, China}
\address{School of Physical Sciences, University of Chinese Academy of Sciences, No.19A Yuquan Road, Beijing 100049, China}

\author{Stefano Chesi}
\address{Beijing Computational Science Research Center, Beijing 100193, China}

\author{Ying-Dan Wang}
\email{yingdan.wang@itp.ac.cn}
\address{Institute of Theoretical Physics, Chinese Academy of Sciences, Beijing 100190, China}
\address{School of Physical Sciences, University of Chinese Academy of Sciences, No.19A Yuquan Road, Beijing 100049, China}
\address{Synergetic Innovation Center for Quantum Effects and Applications, Hunan Normal University, Changsha 410081, China}

\begin{abstract}
We study how strong mechanical squeezing (beyond $3$ dB) can be achieved through reservoir engineering in an optomechanical system which is far from the resolved-sideband regime. In our proposed setup, the effect of unwanted counter-rotating terms is suppressed by quantum interference from two auxiliary cavities. In  the weak coupling regime we develop an analytical treatment based on the effective master equation approach,  which allows us to obtain explicitly the condition of maximum squeezing.
\end{abstract}
\pacs{03.67.Lx, 76.30.Mi, 42.50.Pq, 85.25.Dq}
\maketitle
\end{CJK*}

\section{INTRODUCTION}
Quantum squeezed states of mechanical resonators represent a striking exhibition of macroscopic
quantum effects. Besides their conceptual interest, they have important applications to ultrasensitive measurements and continuous-variable quantum-information processing~\cite{Caves1980rev,Braunstein2005rev}. A standard approach to generate squeezing
is to introduce a coherent drive modulating the mechanical spring constant at twice the mechanical resonance frequency. In cavity optomechanics, such coherent parametric drive can be realized by an amplitude-modulated laser drive~\cite{AMari2009prl,Woolley2008pra,
Nunnenkamp2010pra,JQLiao2011pra,MSchmidt2012njp}. However, due to mechanical instability, the degree of squeezing generated by this approach is bounded by the so-called `$3$ dB limit'~\cite{Milburn1981optc}. In other words, any quadrature cannot
be squeezed below $50\%$ of its zero-point level.

Possible ways to overcome the $3$ dB limit have been proposed, but usually pose significant experimental challenges, e.g., require the assistance of continuous weak measurement and feedback~\cite{Ruskov2005prb,Aashish2008njp,Szorkovszky2011prl,Szorkovszky2013prl}, or a strong intrinsic nonlinearity of the system~\cite{DVitalipra2014,nori2015pra}. Unbounded squeezing can also be
generated by injecting squeezed light into the cavity and transferring optical squeezing into the mechanics~\cite{Jahne2009pra,sumei2010pra}, which requires strong coupling and a highly squeezed broadband field. Instead, a relatively simple way to generate strong mechanical squeezing is based on reservoir engineering~\cite{Kronwaldpra2013}. Such proposal has been demonstrated experimentally \cite{Wollman2015,Pirkkalainen2015,Lecocq2015,Lei2016} and recently the $3$~dB limit has been surpassed in Ref.~\cite{Lei2016}. Nevertheless, all these realizations are based on electromechanical systems, while for optomechanical systems the requirement of achieving the deep resolved-sideband regime is still challenging~\cite{Florian2014rev}. This is mainly due to the difficulty of improving the optical finesse in a cavity with floppy mechanical elements.

\begin{figure}
\begin{centering}
\includegraphics[width=0.4\textwidth,angle=0]{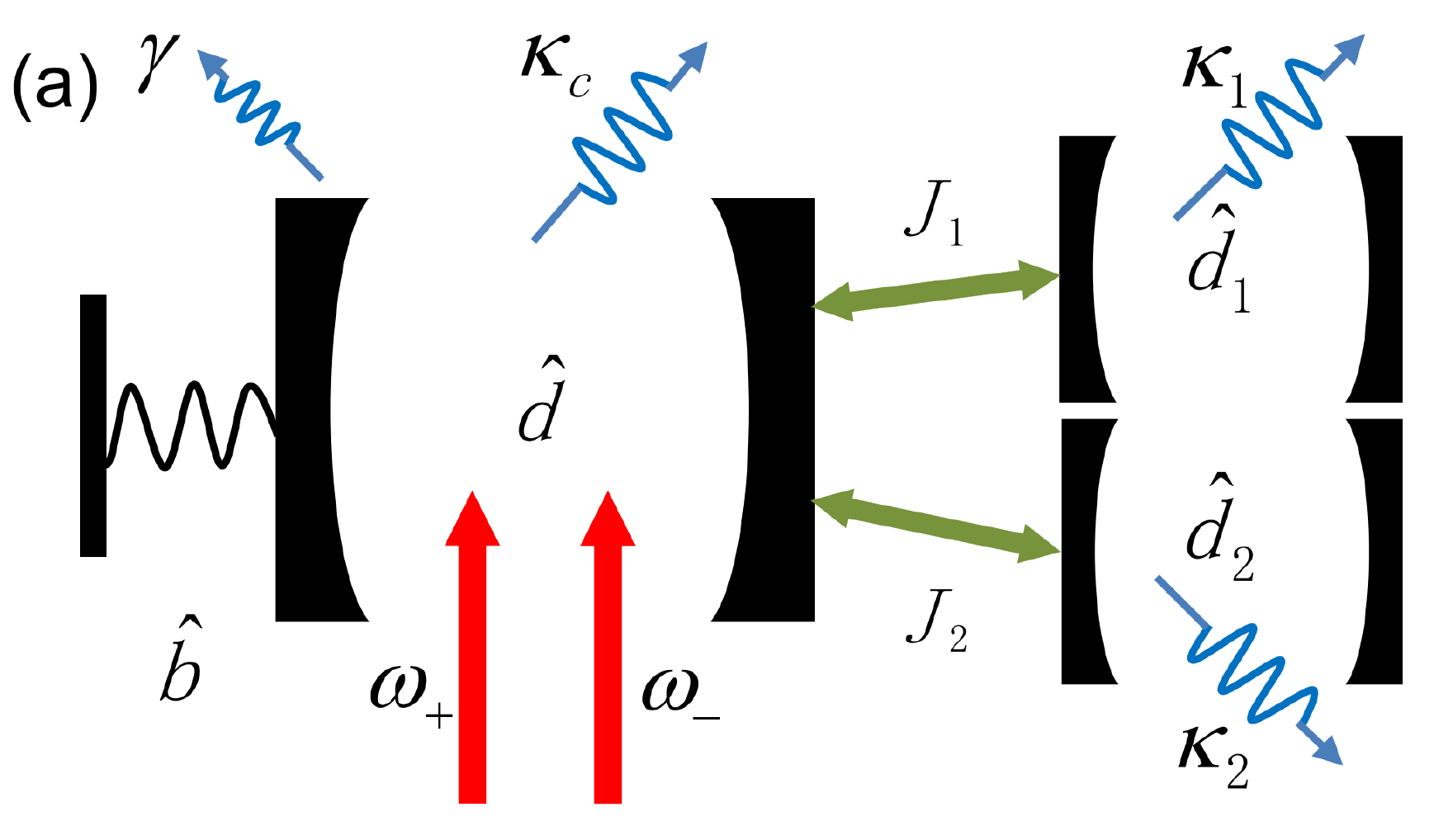}
\includegraphics[width=0.4\textwidth,angle=0]{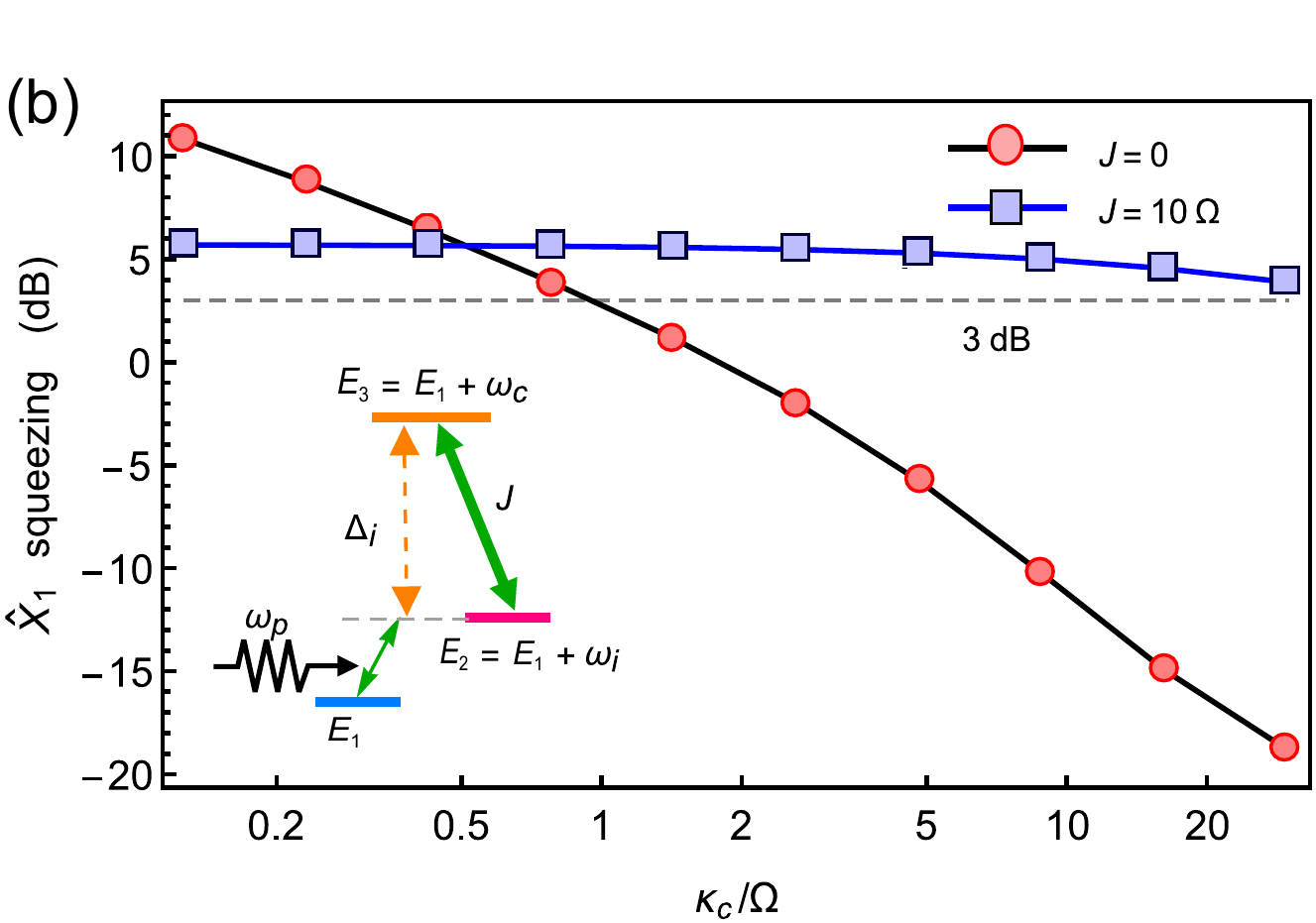}
\end{centering}
\caption{(Color online) (a) System schematics. (b) Mechanical squeezing with/without
the auxiliary cavities: $J/\Omega=0$ (red circles) and $J/\Omega=10$
(blue squares). Both curves have been optimized over $G_{+}$ (at fixed $G_- = \Omega/10$) and a symmetric configuration
is assumed: $J_{1,2}=J$ and $\kappa_{1,2}=\kappa$. Other parameters are:
$\kappa/\Omega=1/2$,
$\Delta_{1,2}/\Omega=\pm2$, $n_{\text{th}}=0$, $\gamma/\Omega=10^{-5}$. Inset:
Partial energy diagram of the coupled optical cavities $\hat{a}$ and $\hat
{a}_{i}$ ($i=1$ or 2). $E_{2}$ ($E_{3}$) is the energy of the state
with one more photon in the $\hat{a}$ mode ($\hat{a}_{i}$ mode). When the
probe light has frequency $\omega_{p}=\omega_{i}$, the two-photon resonance
condition is satisfied and dips appear in the optical spectrum at
those frequencies. In the rotating frame, the dips locate at $\omega_p
=\omega_i-\omega_c = -\Delta_{i}$ [cf. Eq.~(\ref{sop})]. }%
\label{fig_schematics}%
\end{figure}

To address this problem, we propose here an improved version of the reservoir engineering approach. As illustrated in Fig.~\ref{fig_schematics}(a), we consider a driven optomechanical cavity linearly coupled to two auxiliary high-Q cavities (these are pure optical cavities without movable elements and much higher quality factors are realizable). Such linear coupling can be readily implemented in optomechanical systems like microtoroids~\cite{Franco2014prl} or photonic crystal nanobeams~\cite{Susumu2012natpho}. The two auxiliary cavities can be considered as part of the engineered reservoir for the mechanics and, for carefully chosen parameters, provide the fine structure necessary to suppress the two counter-rotating processes in the unresolved-sideband regime. The suppression arises from quantum interference, analogously to electromagnetic induced transparency (EIT), and relies on the coherence properties of the two auxiliary modes. Similar ideas has also been explored in non-resolved sideband cooling~\cite{Ojanenpra2014,yujiepra2014,yongchunpra2015}, with a single auxiliary cavity.

In the following, we first analyze the system based on a full numerical solution of the Langevin equations. As shown in Fig.~\ref{fig_schematics}(b), we find that squeezing beyond $3$ dB in the unresolved-sideband regime can indeed be achieved using this approach, with an appropriate choice of realistic parameters. In the weak-coupling limit, we also derive an effective master equation
for the mechanics, by treating the three coupled optical cavities as an engineered reservoir. Within the effective master equation
approach, we obtain transparent analytical results which allow us to discuss how to maximize squeezing by optimizing system parameters.

The detailed outline is as follows. In Sec.~\ref{sec:model} we briefly review squeezing generation via reservoir engineering, introduce our model, and discuss results obtained by solving the Langevin equations. In Sec.~\ref{sec:weak_coupling} we derive
the weak-coupling effective master equation and the explicit expression for the the steady-state mechanical variance.
In Sec.~\ref{sec:spectrum} we analyze the spectrum of the coupled optical cavities.
In Sec.~\ref{sec:optimiztion} we derive the conditions to achieve maximum squeezing and discuss the experimental feasibility. Section~\ref{sec:imperfections} contains our concluding remarks and Appendices~\ref{app_linearization}-\ref{app_dependence} discuss some technical details.


\section{MODEL}\label{sec:model}

We consider the system schematically shown in Fig.~\ref{fig_schematics}(a) and
described by the following Hamiltonian:
\begin{align}
\hat{H}=  &  \omega_{c}\hat{a}^{\dag}\hat{a}+\Omega\hat{b}^{\dagger}\hat
{b}+g\hat{a}^{\dagger}\hat{a}\left(  \hat{b}^{\dagger}+\hat{b}\right)
+\hat{H}_{\text{dr}}\nonumber\\
&  +\sum_{i=1,2}\left(  \omega_{i}\hat{a}_{i}^{\dagger}\hat{a}_{i}+J_{i}%
(\hat{a}^{\dagger}\hat{a}_{i}+\hat{a}\hat{a}_{i}^{\dagger})\right)  + \hat
{H}_{\mathrm{env}}, \label{OH}%
\end{align}
where $\hat{a}$ is the annihilation operator of the main cavity (frequency $\omega_c$), which is
coupled to a mechanical mode (annihilation operator $\hat{b}$, frequency $\Omega$) and two auxiliary
cavities (annihilation operators $\hat{a}_{1,2}$, frequencies $\omega_{1,2}$). The coupling to the mechanical mode is a standard
optomechanical interaction with coupling strength $g$. $J_{1,2}$ are the coupling constants between
the main and auxiliary cavities. To induce squeezing of
the mechanical state, a two-tone drive is applied to the main cavity~\cite{Kronwaldpra2013,schwabsci2017}:
\begin{equation}
\hat{H}_{\text{dr}}=\left(  \alpha_{+}e^{-i\omega_{+}t}+\alpha_{-}%
e^{-i\omega_{-}t}\right)  \hat{a}^{\dagger}+\text{H.c.},
\end{equation}
where $\omega_{\pm}=\omega_{c}\pm\Omega$ are the frequencies of the two laser drives. Finally, $\hat{H}_{\mathrm{env}}$ describes the coupling to
Markovian reservoirs. As indicated in Fig.~\ref{fig_schematics}(a), the
damping rates of the main cavity, auxiliary cavities, and mechanics are
respectively given by $\kappa_{c}$, $\kappa_{1,2}$, and $\gamma$.

For a weak optomechanical interaction $g<\kappa,\gamma$, the Hamiltonian
can be linearized with the standard procedure, where we perform the displacement transformations
$\hat{a}=\alpha+\hat{d}$ for the main cavity, $\hat{a}_{i}=\alpha_{i}+\hat{d}_{i}$ for
the auxiliary cavities ($i=1,2$), and $\hat{b}=\beta+\hat{d}_{\mathrm{m}}$ for the mechanical mode,
and neglect small nonlinear effect (see details in
Appendix~\ref{app_linearization}). Finally, in a suitable rotating frame, the linearized
Hamiltonian reads:
\begin{align}
\label{formula_ILH}\hat{H}_{I}=  &  \hat{d}^{\dag}(G_{+}\hat{d}_{\mathrm{m}%
}^{\dagger}+G_{-}\hat{d}_{\mathrm{m}})+\text{H.c.}\nonumber\\
&  +\hat{d}^{\dag}(G_{+}\hat{d}_{\mathrm{m}}e^{-2i\Omega t}+G_{-}\hat
{d}_{\mathrm{m}}^{\dagger}e^{2i\Omega t})+\text{H.c.}\nonumber\\
&  +\sum_{i=1,2}\left(  -\Delta_{i}\hat{d}_{i}^{\dag}\hat{d}_{i}+J_{i}(\hat
{d}^{\dag}\hat{d}_{i}+\hat{d}_{i}^{\dag}\hat{d})\right)  + \hat{H}%
_{\mathrm{env}} ,
\end{align}
where $\Delta_{1,2}=\omega_{c}-\omega_{1,2}$ and $G_{\pm}$
are the dressed optomechanical couplings.
The first line of Eq.~(\ref{formula_ILH}) realizes the
standard squeezing via reservoir engineering~\cite{Kronwaldpra2013}, since the
cavity can cool the mechanical Bogoliubov mode
\begin{equation}
\label{Bmode}\hat{B}\equiv\hat{d}_{\mathrm{m}
}\cosh\zeta-\hat{d}_{\mathrm{m}}^{\dag}\sinh\zeta,
\end{equation}
where the squeezing parameter is $\tanh\zeta=G_{+}/G_{-}$.
As the vacuum of $\hat{B}$ is exactly the mechanical squeezed state
$\left\vert 0\right\rangle _{\hat{B}}=\exp[\zeta(\hat{d}_{\mathrm{m}}^{2}%
-\hat{d}_{\mathrm{m}}^{\dag2})/2]\left\vert 0\right\rangle _{\hat
{d}_{\mathrm{m}}}$, cooling of mode $\hat{B}$ directly yields mechanical
squeezing. Note that the coefficients of the Bogoliubov transformation are
real, thus the maximally squeezed quadrature is $\hat{X}_{1}=( \hat
{d}_{\mathrm{m}}+\hat{d}_{\mathrm{m}}^{\dag})/\sqrt{2}$ (see
Appendix~\ref{appendix_maxsqueezing}), with variance $e^{-2\zeta}/2$.

Such an ideal cooling of the $\hat{B}$ mode becomes impossible in the non-resolved
sideband regime ($\kappa_{c}>\Omega$), as one cannot neglect the two
counterrotating terms appearing in the second line of Eq.~(\ref{formula_ILH}).
With respect to the original mechanical mode, the first counterrotating term ($\propto G_{+}$, induced by the upper sideband laser
drive) has a cooling effect on $\hat{d}_{\mathrm{m}}$, while the second
counterrotating term ($\propto G_{-}$, induced by the lower sideband laser
drive) has a heating effect on $\hat{d}_{\mathrm{m}}$. Both processes lead to heating of the Bogoliubov mode $\hat{B}$. Due to
the large optical state density at these frequencies, mechanical squeezing cannot be achieved in
the unresolved sideband regime. The degradation of squeezing with $\kappa_{c}$ is illustrated in
Fig.~\ref{fig_schematics}(b) where the squeezing is quantified through:
\begin{equation}
S_{\mathrm{dB}}=-10 \log_{10} \left[  2\langle\Delta\hat{X}_{1}^{2}%
\rangle\right]  .
\end{equation}
As seen from the plot, the maximum achievable
squeezing decreases with the increasing cavity damping and large squeezing is only achievable in
the resolved sideband regime. A quantitative analysis regarding this point
will be given in Section V.

Figure~\ref{fig_schematics}(b) also shows that turning on the
couplings with the auxiliary cavities can greatly improve the performance when $\kappa_c/\Omega >1$. Even in the bad cavity limit
($\kappa_{c}/\Omega=10$), squeezing beyond $3$ dB is achievable under appropriate conditions,
which will be discussed in the rest of the paper.
The general principle is that the auxiliary cavities allow us to modulate the optical
density of states through destructive interference, and therefore alleviate the damaging effects of the
counter-rotating terms.

The two curves of Fig.~\ref{fig_schematics}(b) are obtained by numerically
solving the Langevin equations of the full system~\cite{yingdan2012njp}.
In the following, to gain physical understanding of the mechanism, we pursue an approach
based on the effective master equation for the mechanical mode. This treatment is valid in the weak-coupling
regime and provides explicit analytical expressions for the optimal working point and
maximum squeezing.

\section{Mechanical squeezing in the weak-coupling regime}\label{sec:weak_coupling}
At weak coupling, i.e., $G_{\pm}\ll\kappa_{c},\kappa_{1,2}$, the
interacting cavities can be viewed as a structured environment for
the mechanics. Hence, as described with more detail in
Appendix~\ref{appendix_ME}, we can follow the standard Born-Markov
procedure~\cite{opensystem} and trace out the cavity degrees of freedom. As a
result, we obtain the following effective master equation for the mechanics:
\begin{align}
\frac{d\hat{\rho}\left( t\right)  }{dt}=  &  \Gamma_{-}\mathcal{D}(\hat
{d}_{\mathrm{m}})\hat{\rho}+\Gamma_{+}\mathcal{D}(\hat{d}_{\mathrm{m}}^{\dag
})\hat{\rho}\nonumber\\
&  +\Gamma_{S}\left(\mathcal{D}_{S}(\hat{d}_{\mathrm{m}})\hat{\rho
}+\mathcal{D}_{S}(\hat{d}_{\mathrm{m}}^{\dag})\hat{\rho}\right)  .
\label{formula_master}%
\end{align}
Here $\mathcal{D}(\hat{A})\hat{\rho}=\hat{A}\hat{\rho}\hat{A}^{\dag}-\frac
{1}{2}\hat{A}^{\dag}\hat{A}\hat{\rho}-\frac{1}{2}\hat{\rho}\hat{A}^{\dag}%
\hat{A}$ is a standard dissipator, thus $\mathcal{D}(\hat{d}_{\mathrm{m}})$
and $\mathcal{D}(\hat{d}^{\dag}_{\mathrm{m}})$ represent cooling and heating effects caused
by the optical cavities and thermal environment. The corresponding rates are
given by:
\begin{align}
\label{damp}\Gamma_{-}  &  =\gamma\left(  1+n_{\text{th}}\right)  + G_{-}%
^{2}S_{\text{op}}\left(  0\right)  +G_{+}^{2}S_{\text{op}}\left(
2\Omega\right)  ,\nonumber\\
\Gamma_{+}  &  =\gamma n_{\text{th}} + G_{+}^{2}S_{\text{op}}\left(  0\right)
+G_{-}^{2}S_{\text{op}}\left(  -2\Omega\right)  ,
\end{align}
with the optical spectral function:
\begin{equation}
S_{\text{op}}(\omega)=\int_{-\infty}^{+\infty}dte^{i\omega t}\langle\hat
{d}(t)\hat{d}^{\dag}(0)\rangle,
\end{equation}
which will be extensively discussed in the next section. Here we only note that $S_{\text{op}}(\omega)$ is a real quantity, which can be easily shown using $\langle\hat{d}(t)\hat{d}^{\dag}(0)\rangle = \langle\hat {d}(0)\hat{d}^{\dag}(-t)\rangle$.

Equation~({\ref{damp}}) shows how the
standard mechanical dissipation, given in terms of damping $\gamma$ and
thermal occupation $n_{\mathrm{th}}$, can be strongly modified by the optical
environment. In particular, $G_{\pm}^{2}S_{\text{op}}\left(  0\right)  $
is contributed from the rotating-wave terms, while $G_{+}^{2}S_{\text{op}}\left(  2\Omega\right)  $
and $G_{-}^{2}S_{\text{op}}\left(  -2\Omega\right)  $ originate from the
counter-rotating terms in the Hamiltonian Eq.~(\ref{formula_ILH}). In the case of
resolved sideband, only $G_{\pm}^{2}S_{\text{op}}\left(  0\right)$ contributes significantly.

While the first line of Eq.~(\ref{formula_master}) would simply lead to a thermal state of the mechanical mode, the stationary solution is modified by the squeezing superoperators in the second line. They are given by $\mathcal{D}_S(\hat{A})\hat{\rho}=\hat{A}\hat{\rho}\hat{A}-\frac{1}{2}\hat{A}\hat{A}\hat{\rho}-\frac{1}{2}\hat{\rho}\hat{A}\hat{A}$, with the rate:
\begin{equation}
\Gamma_{S}=G_{+}G_{-}S_{\text{op}}\left(  0\right) . \label{squeeze}%
\end{equation}
The generation of squeezing can be ascribed to the presence of such terms.

\subsection{General formula for the squeezed quadrature}

The master equation becomes physically more transparent when rewritten in Lindblad form.
Equation~(\ref{formula_master}) deviates from the Lindblad form due to the squeezing terms $\mathcal{D}_{S}(\hat{d}_{\rm m})$ and
$\mathcal{D}_{S}(\hat{d}_{\rm m}^{\dag})$, whose role is to induce squeezing by relaxing the mechanics to a thermal state of a certain Bogoliubov mode.

For example, in the extreme resolved sideband limit  we have $S_{\text{op}}\left(  0\right)
=4/\kappa_{c}$ and $S_{\text{op}}\left(  \pm2\Omega\right)  =0$. Neglecting the small mechanical damping
$\gamma=0$, Eq.~(\ref{formula_master}) reads:
\begin{equation}
\frac{d\hat{\rho}\left(  t\right)  }{dt}=\Gamma_{\text{opt}}\mathcal{D}%
(\hat{B})\hat{\rho}, \label{MEQ_B}%
\end{equation}
where $\Gamma_{\text{opt}}=4(G_{-}^{2}-G_{+}^{2})/\kappa_{c}$. This limit is
in agreement with our previous discussion about relaxation into the vacuum of
the $\hat{B}$ mode.

In the general case, The Lindblad form of Eq.~(\ref{formula_master}) is derived as follows (see details in Appendix~\ref{appendix_DME}):
\begin{equation}
\frac{d\hat{\rho}\left(  t\right)  }{dt}=\Gamma^{B'}_{-}\mathcal{D}(  \hat
{B}')  \hat{\rho}+\Gamma^{B'}_{+}\mathcal{D}(  \hat{B}'^{\dagger})  \hat{\rho},
\label{MEQ_cmode}%
\end{equation}
where the new Bogolubov mode is
\begin{equation}
\hat{B}'\equiv\frac{\Gamma_{S}}{b}\sqrt{\frac{2b}{a-b}}\hat{d}_{\mathrm{m}}%
+\sqrt{\frac{a-b}{2b}}\hat{d}_{\mathrm{m}}^{\dagger}, \label{mode_c}%
\end{equation}
with $a=\Gamma_{-}+\Gamma_{+}$  and $b=\sqrt{(\Gamma_{-}+\Gamma_{+})^{2}-4\Gamma_{S}^{2}}$. The corresponding rates are
$\Gamma^{B'}_{\pm}=(\Gamma_{\pm}-\Gamma_{\mp}+b)/2$. Setting $\Gamma^{B'}_{-}> \Gamma^{B'}_{+}$ (or, equivalently $\Gamma_- > \Gamma_+$) we obtain the stability condition:
\begin{equation}
\frac{G_+^2}{G_-^2} < \frac{1-\varepsilon_-+1/C_{\rm e}}{1-\varepsilon_+}, \label{stability}
\end{equation}
where we defined the effective cooperativity
\begin{equation}
C_{\text{e}}=G_{-}^{2}S_{\text{op}}\left(  0\right)  /\gamma \label{Ce}
\end{equation}
and the parameters $\varepsilon_{\pm}$, characterizing the strength of the counter-rotating terms:
\begin{equation}
\varepsilon_{\pm}=S_{\text{op}}(\pm2\Omega)/S_{\text{op}}(0). \label{eps_pm}
\end{equation}

The stationary state of Eq.~(\ref{MEQ_cmode}) is a thermal state of mode $\hat{B}'$
and, since the coefficients in Eq.~(\ref{mode_c}) are real, the largest squeezing is obtained
for the $\hat{X}_{1}$ quadrature. The final result reads:
\begin{align}
&  \left\langle \Delta\hat{X}_{1}^{2}\right\rangle =\left\langle \hat{X}%
_{1}^{2}\right\rangle -\left\langle \hat{X}_{1}\right\rangle ^{2}\nonumber\\
&  =\frac{1}{2}\frac{e^{-2\zeta}+\left(  \varepsilon_{-}+\left(
1+2n_{\text{th}}\right)  /C_{\text{e}}\right)  \cosh^{2}\zeta+\varepsilon
_{+}\sinh^{2}\zeta}{1+\left(  1/C_{\text{e}}-\varepsilon_{-}\right)  \cosh
^{2}\zeta+\varepsilon_{+}\sinh^{2}\zeta}, \label{svariance1}%
\end{align}%
where the denominator is always positive, due to the stability condition Eq.~(\ref{stability}).
Equation~(\ref{svariance1}) shows how the ideal squeezing $e^{-2\zeta}/2$ of Eqs.~(\ref{Bmode}) and (\ref{MEQ_B}) is degraded by the effect of counter-rotating terms (giving $\varepsilon_{\pm}\neq0$) and mechanical damping (giving $1/C_{\mathrm{e}}\neq0$).
Intuitively speaking, stronger squeezing requires larger $C_{\text{e}}$ and smaller $\varepsilon_\pm$, and this is also easy to show from analyse of Eq.~(\ref{svariance1}) (see Appendix~\ref{app_dependence}). However, in the bad cavity regime and without coupling to the auxiliary cavities, $S_{\text{op}}(\pm2\Omega)$ is comparable to $S_{\text{op}}(0)$ and the relatively large value of $\varepsilon_{\pm}$ (reflecting significant heating of mode $\hat{B}$) degrades the mechanical squeezing, see Fig.~\ref{fig_schematics}(b). Quantum interference in the coupled cavity system allows to decrease $S_{\text{op}}(\pm2\Omega)$ and achieve squeezing beyond 3~dB. In the next section we will discuss in detail how to modulate the optical spectrum to achieve this goal.

\section{SPECTRUM OF THE STRUCTURED ENVIRONMENT}\label{sec:spectrum}

From the above discussion, we see that the values of the optical spectrum $S_{\rm op}(\omega)$
at $\omega=0,\pm 2\Omega$ are crucial to achieve strong mechanical squeezing. In
the following, we investigate the dependence of the optical spectrum on system
parameters and how to set up the cavities to achieve strong squeezing.

In the weak coupling regime, the back-action of mechanics to the optical
cavities can be neglected, thus the optical spectrum is determined by the Hamiltonian:
\begin{equation}
\hat{H}_{\text{BO}}=\sum_{i=1,2}\left[  -\Delta_{i}\hat{d}_{i}^{\dag}\hat
{d}_{i}+J_{i}(\hat{d}_{i}^{\dag}\hat{d}+\hat{d}^{\dag}\hat{d}_{i})\right] + \hat{H}^O_{\rm env},
\label{HBO}%
\end{equation}
where $\hat{H}^O_{\rm env}$ describes the baths of the optical cavities. The corresponding quantum Langevin equations are:
\begin{align}
\dot{\hat{d}} &  =-iJ_{1}\hat{d}_{1}-iJ_{2}\hat{d}_{2}-
\frac{\kappa_{c}}{2}\hat{d}+\sqrt{\kappa_{c}}\hat{d}_{c,{\rm in}}, \label{LE_c}\\
\dot{\hat{d}}_{i}  &  =-iJ_{i}\hat{d}+i\Delta_{i}\hat{d}_{i}%
-\frac{\kappa_{i}}{2}\hat{d}_{i}+\sqrt{\kappa_{i}}\hat{d}_{i,{\rm in}}, \label{LE_i}
\end{align}
where in Eq.~(\ref{LE_i}) $i=1,2$ and the noise operators $\hat{d}_{\alpha,{\rm in}}$ ($\alpha =c,1,2$) satisfy $\langle\hat{d}_{\alpha,{\rm in}}(t)\hat{d}_{\beta,{\rm in}}^{\dag}(t')\rangle=\delta_{\alpha\beta}\delta(t-t')$. The above Langevin equations yield the following spectrum:
\begin{equation}\label{sop}
S_{\text{op}}( \omega)  =\frac{1}{A\left(  \omega\right)  }%
+\frac{1}{A^{\ast}\left(  \omega\right)  },
\end{equation}
with
\begin{equation}\label{Adef}
A\left(  \omega\right)  =\frac{\kappa_{c}}{2}-i\omega+i\sum_{j=1,2}\frac{J_{j}^{2}}%
{ \omega+\Delta_{j}+i\kappa_{j}/2 }.
\end{equation}

\begin{figure}
 \begin{centering}
    \includegraphics[width=0.45\textwidth,angle=0]{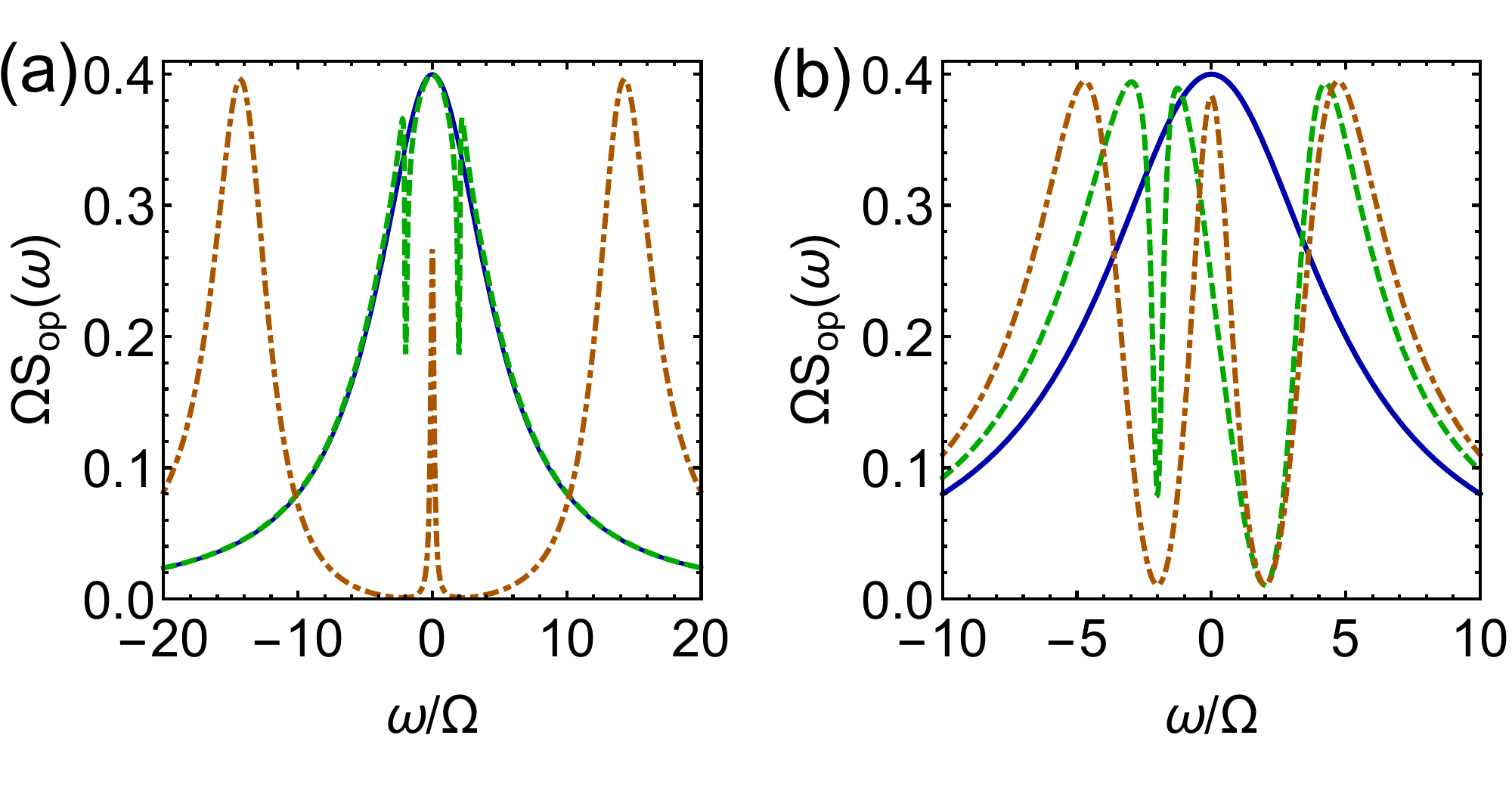}
    \end{centering}
    \caption{(Color online) (a) Optical spectrum for small $J/\Omega=1/2$
    (green dashed), large $J/\Omega=10$ (orange dashed dot), and without auxiliary
    cavities (blue). (b) Optical spectrum with different couplings, $J_{1}/\Omega=1$ and $J_{2}/\Omega=3$ (green dashed). As a reference, we also plot the spectrum with $J_{1}/\Omega=J_{2}/\Omega=3$ (red dot-dashed) and $J_{1}/\Omega=J_{2}/\Omega=0$ (blue). In both panels, the decay of the main cavity and the auxiliary cavities are $\kappa
    _{c}/\Omega=10$ and $\kappa/\Omega=1/10$, respectively.}%
    \label{spectrum}%
    \end{figure}

Some representative plots of $S_{\text{op}}( \omega)$ are shown in Fig.~\ref{spectrum}.
Without auxiliary cavities, the optical spectrum has a Lorentzian shape
with a single peak located at $\omega=0$, the width of the peak being
$\kappa_{c}$. In the deep unresolved-sideband regime $\kappa_c \gg \Omega$, the values of $S_{\text{op}%
}\left( \pm2\Omega\right)  $ are close to $S_{\text{op}}\left(
0\right)  $ (i.e. $\varepsilon_\pm\approx 1$), and the mechanical squeezing effect is suppressed.

With two coupled cavities, the simple Lorentzian line shape is modified. Two dips emerge
at $-\Delta_{1}$ and $-\Delta_{2}$, i.e., at the two-photon resonance
condition ($E_3-E_1-\omega_p=E_3-E_2$ see inset of Fig.~\ref{fig_schematics}(b)). Furthermore, the position of the central peak remains unchanged
if $\kappa_1=\kappa_2$ and $J_1=J_2$. To achieve small values of $\varepsilon_{\pm}$,
$S_{\text{op}}(\pm2\Omega)$ should be minimized while $S_{\text{op}}\left(  0\right) $ should
be maximized. Hence, a natural choice is to set the two dips at frequency $\pm 2\Omega$ and the peak at frequency $0$,
i.e., $\Delta_{1}=-\Delta_{2}=2\Omega$, $J_{1}=J_{2}=J$, and
$\kappa_{1}=\kappa_{2}=\kappa$ (some effects of asymmetry will be discussed in Sec.~\ref{sec:imperfections}). With this symmetric setting, $\varepsilon_{-}=\varepsilon_{+}=\varepsilon$ and
\begin{equation}
S_{\text{op}}\left(  0\right)  =\frac{2}{\kappa_{c}/2+J^{2}\kappa/\left(
\kappa^{2}/4+4\Omega^{2}\right)  }. \label{S0}
\end{equation}
When $J,\kappa\ll\Omega$, this expression reduces to the result without the auxiliary cavities $S_{\text{op}}\left(  0\right)
\approx4/\kappa_{c}$.

In the large $J$ limit ($J\gg\kappa_c$), which is analogous to the
Autler-Townes regime, we find three distinct resonances located at
$\omega=\pm\sqrt{2J^{2}+4\Omega^{2}}$ and $0$, obtained by diagonalizing
$\hat{H}_{\mathrm{BO}}$.
The width of the middle peak is $\left(  \kappa_{c}-\kappa\right) \Omega^2 /J^{2}+\kappa/2$, i.e., is limited by the linewidth of the auxiliary cavities, with its height suppressed by the coupling $J$ [see Eq.~(\ref{S0})]. The width of the two side peaks is $\left(  \kappa_{c}+\kappa\right)
/4-\left(  \kappa_{c}-\kappa\right)\Omega^2  /\left(  2J^{2}\right)  $.
For small $J\ll\kappa_c$, the optical spectrum follows a lineshape similar to EIT,
with two narrow dips at $\omega=\pm 2\Omega$ (cf. Fig.~\ref{spectrum}(a)). However, for typical parameters of this system, we find that
the optimal $J$ should be on the same order of $\kappa_c$ (see Sec.~\ref{sec:optimal_J}). Then, the spectrum takes an intermediate shape of the type shown in Fig.~\ref{spectrum}(b).

To characterize the dependence of $\varepsilon$, we should consider $S_{\text{op}}\left(  \pm 2\Omega\right)$. If the auxiliary cavities are weakly damped, such that $\kappa\ll\Omega$, and assuming $J^{2}\gg\kappa_{c}\kappa$, one has:
\begin{equation}
S_{\text{op}}\left(  \pm 2\Omega\right)  \approx\frac{\kappa}{J^{2}}.
\label{Sop1}%
\end{equation}
Then, Eqs.~(\ref{S0}) and (\ref{Sop1}) lead to:
\begin{equation}
\varepsilon\simeq\frac{\kappa_{c}\kappa}{4J^{2}}+\frac{\kappa^{2}}{8\Omega
^{2}}, \qquad(\kappa\ll\Omega) \label{epsilon_largeJ}
\end{equation}
which is a decreasing function of $J$ and saturates to the lower bound $\varepsilon\simeq\kappa^{2}/(8\Omega^{2})$
when $J^{2}\gg(\Omega/\kappa)^{2}\kappa_{c}\kappa$. Note that Eq.~(\ref{epsilon_largeJ}) is also a decreasing function of the ratio $\kappa/\Omega$. In
conclusion, to decrease the value of $\varepsilon$, it is beneficial to set $\Delta_{1}=-\Delta_{2}=2\Omega$, increase $J$, and decrease $\kappa$. At the same time, it is important to note that a larger $J$ suppresses the effective cooperativty $C_{\rm e}$.

\section{OPTIMIZITION OF THE MECHANICAL SQUEEZING}\label{sec:optimiztion}

So far, we have discussed the desirable setting of the auxiliary cavities. In this section, we focus
on how to achieve the maximum squeezing effect by optimizing the coupling strength of the main optomechanical cell to the drives and to the auxiliary cavities.

\subsection{Optimal mechanical squeezing with respect to laser strength}\label{sec:optimal_r}

\begin{figure}
\begin{centering}
\includegraphics[width=0.4\textwidth,angle=0]{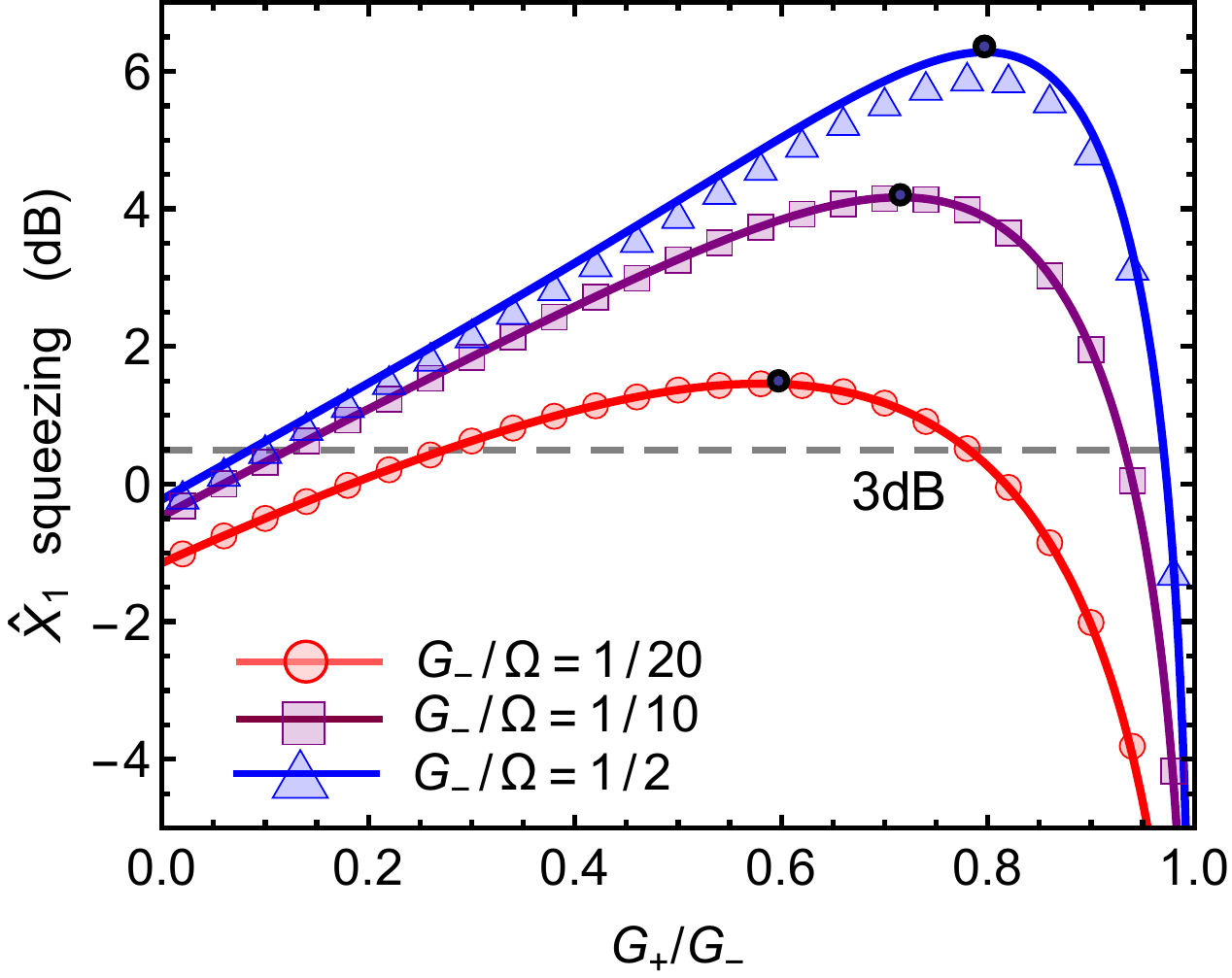}
\par\end{centering}
\caption{(Color online) Mechanical squeezing versus $G_{+}/G_{-}$, with different values of
$G_{-}$. The solid curves are from Eq.~(\ref{svariance1}), while the discrete points (circles, squares, and triangles) are from a numerical solution of the Langevin equations. As expected, small deviations appear at the largest $G_-$ (approaching strong-coupling). The black dots marks the approximated optimal points, i.e., Eqs.~(\ref{al}) and (\ref{avl}). We used $J/\Omega =5$, $\kappa/\Omega =1/5$, $\kappa_{c}/\Omega =10$, and $\gamma /\Omega =10^{-5}$. The thermal bath population is $n_{th}=10$. }%
\label{variance}%
\end{figure}

With the optical parameters of the auxiliary cavities fixed as in the previous section, the mechanical
squeezing effect varies with respect to the strength of the applied lasers. Especially, it can be rather sensitive to the relative strength of the blue- and red-detuned drives, which we define as $r\equiv \tanh\zeta= G_{+}/G_{-}$.

In Fig.~\ref{variance}, the variance $\langle \Delta\hat{X}_{1}^{2}\rangle$ is plotted as function of $r$ for several values of $G_-$. Like in the resolved-sideband regime (i.e., without auxiliary cavities), the squeezing has a maximum with respect to $r$. By increasing $r$, the squeezing parameter becomes larger, but at the same time the influence of counter-rotating terms and heating is also enhanced~\cite{Kronwaldpra2013,YDWang2013prl}. A balance between these two opposite effects leads to an optimal value of $r$. For fixed $G_{-}$ and $\varepsilon<1$, this optimal value can be derived from Eq.~(\ref{svariance1}):
\begin{equation}
r_{\text{opt}}=\frac{D - \sqrt{D
^{2}-C_{\text{e}}(1-\varepsilon)\left(  C_{\text{e}}\left(  1- \varepsilon\right)  +1\right)
}}{C_{\text{e}}\left(  1-\varepsilon\right)  }. \label{OR}%
\end{equation}
where
\begin{equation}
D=C_{\text{e}}(1-\varepsilon^{2})+n_{\text{th}}(1-\varepsilon)+1.%
\end{equation}
The corresponding optimal mechanical variance is:%
\begin{equation}
\left\langle \Delta\hat{X}_{1}^{2}\right\rangle _{r_\text{opt}}%
=\frac{1+2n_{\text{th}}+C_{\text{e}}\left(  \left( r_\text{opt}-1\right)  ^{2}%
+\varepsilon\left(  r_\text{opt}^{2}+1\right)  \right)  }{2\left(1+C_{\text{e}}\left(
\varepsilon-1\right)  \left(  r_\text{opt}^{2}-1\right) \right)}. \label{ORR}%
\end{equation}

\begin{figure}
\begin{centering}
\includegraphics[width=0.4\textwidth,angle=0]{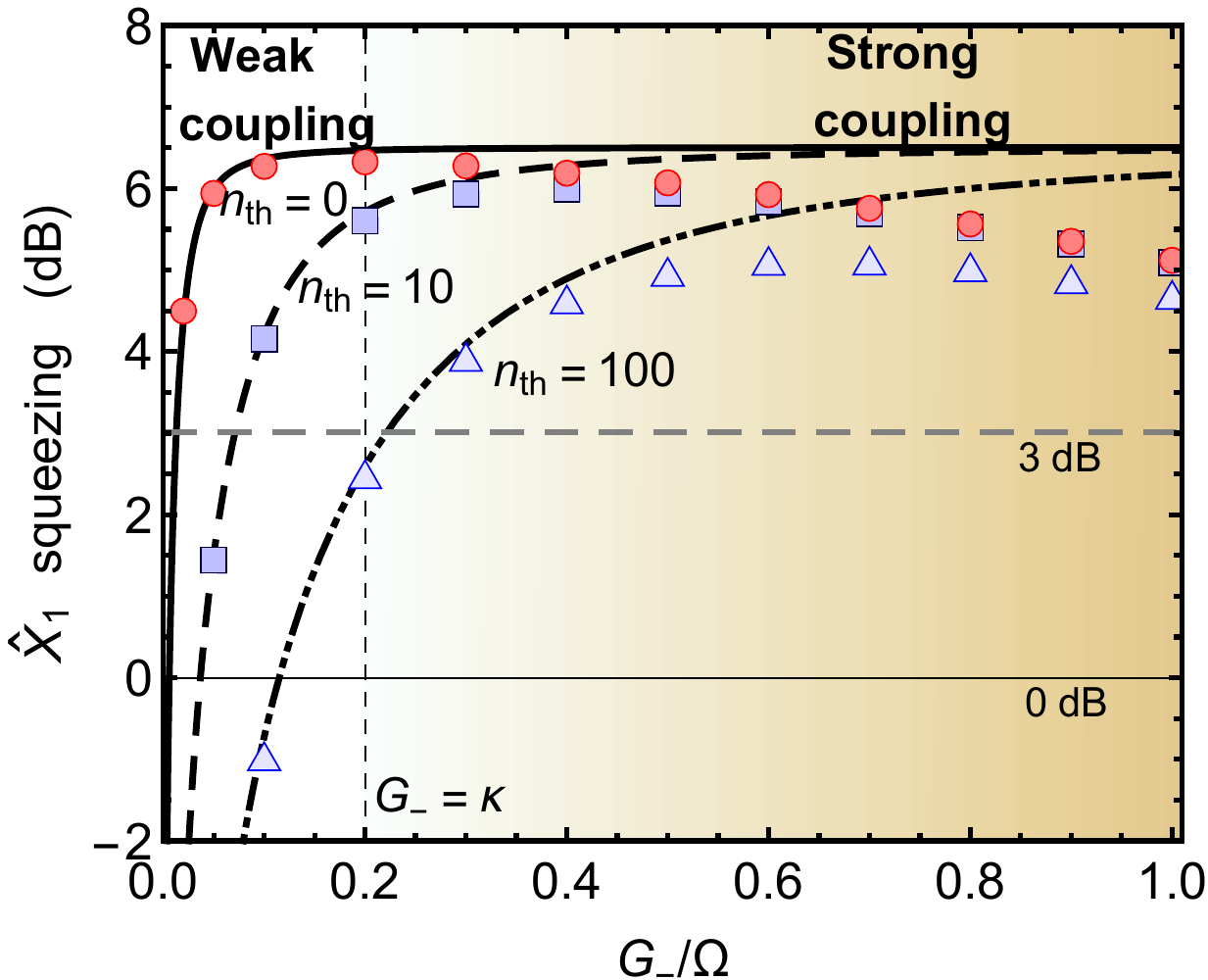}
\par\end{centering}
\caption{(Color online) Mechanical squeezing with optimized $r$. All lines are the analytical results from
Eqs.~(\ref{al}) and (\ref{avl}) while circles, squares, and triangles are the corresponding numerical results. The boundary between the weak- and strong-coupling regime is marked by $G_{-}=\kappa$. In the strong coupling regime, the effective master
equation approach becomes inadequate. The parameters used here are $J/\Omega =5$, $\kappa /\Omega =1/5$, $\kappa_{c}/\Omega =10$, and $\gamma /\Omega =10^{-5}.$}\label{optimal}
\end{figure}

Considering the relevant limit of large effective cooperativity $C_{\text{e}}\gg 1$ and small
counter-rotating effect $\varepsilon\ll 1$, Eq.~(\ref{OR}) can be simplified to:
\begin{equation}
r_{\text{opt}}\approx1-\sqrt{\frac{1+2C_{\text{e}%
}\varepsilon+2n_{\text{th}}}{C_{\text{e}}}}+\frac{1+C_{\text{e}}%
\varepsilon+n_{\text{th}}}{C_{\text{e}}}. \label{al}%
\end{equation}
In this regime, the minumum variance is:%
\begin{equation}
\left\langle \Delta\hat{X}_{1}^{2}\right\rangle_{r_{\text{opt}}}\approx
\sqrt{\frac{1+2C_{\text{e}}\varepsilon+2n_{\text{th}}}{4C_{\text{e}}}}%
+\frac{n_{\text{th}}}{2C_{\text{e}}}. \label{avl}%
\end{equation}
Figure~\ref{optimal} shows a comparison of the above Eqs.~(\ref{al}) and (\ref{avl}) with the numerical results. From  Eq.~(\ref{avl}) we see that the variance decreases monotonically with $G_{-}$ and saturates at:
\begin{equation}
\left\langle \Delta\hat{X}_{1}^{2}\right\rangle |_{\text{bound}}=\sqrt{\varepsilon/2}\approx\sqrt
{\frac{\kappa_c\kappa}{8 J^2}+\frac{\kappa^{2}}{16\Omega^{2}}}, \label{bound1}
\end{equation}
where in the last step we used Eq.~(\ref{epsilon_largeJ}). This lower bound implies $\langle \Delta\hat{X}_{1}^{2}\rangle >\kappa/(4\Omega)$, which shows that squeezing beyond $3$~dB requires $\kappa<\Omega$.

In Fig.~\ref{variance}, small deviations between our analytical results and the direct numerical solution are visible when $G_-$ is large, due to the violation of the weak-coupling condition. This issue is explored more systematically in Fig.~\ref{optimal}, where the optimal mechanical variance is plotted with respect to $G_{-}$. In the weak-coupling regime, the analytical results are consistent with numerical results. In the strong-coupling regime, the numerical results deviate from the analytical ones,
showing a nonmonotonic behavior with respect of $G_{-}$. This is due to the significant hybridization of the mechanical and optical modes in the strong coupling regime, which invalidates the whole reservoir engineering approach towards a mechanical squeezed vacuum.

\subsection{Optimal mechanical squeezing with respect to $J$}\label{sec:optimal_J}

\begin{figure}
\begin{centering}
\includegraphics[width=0.45\textwidth,angle=0]{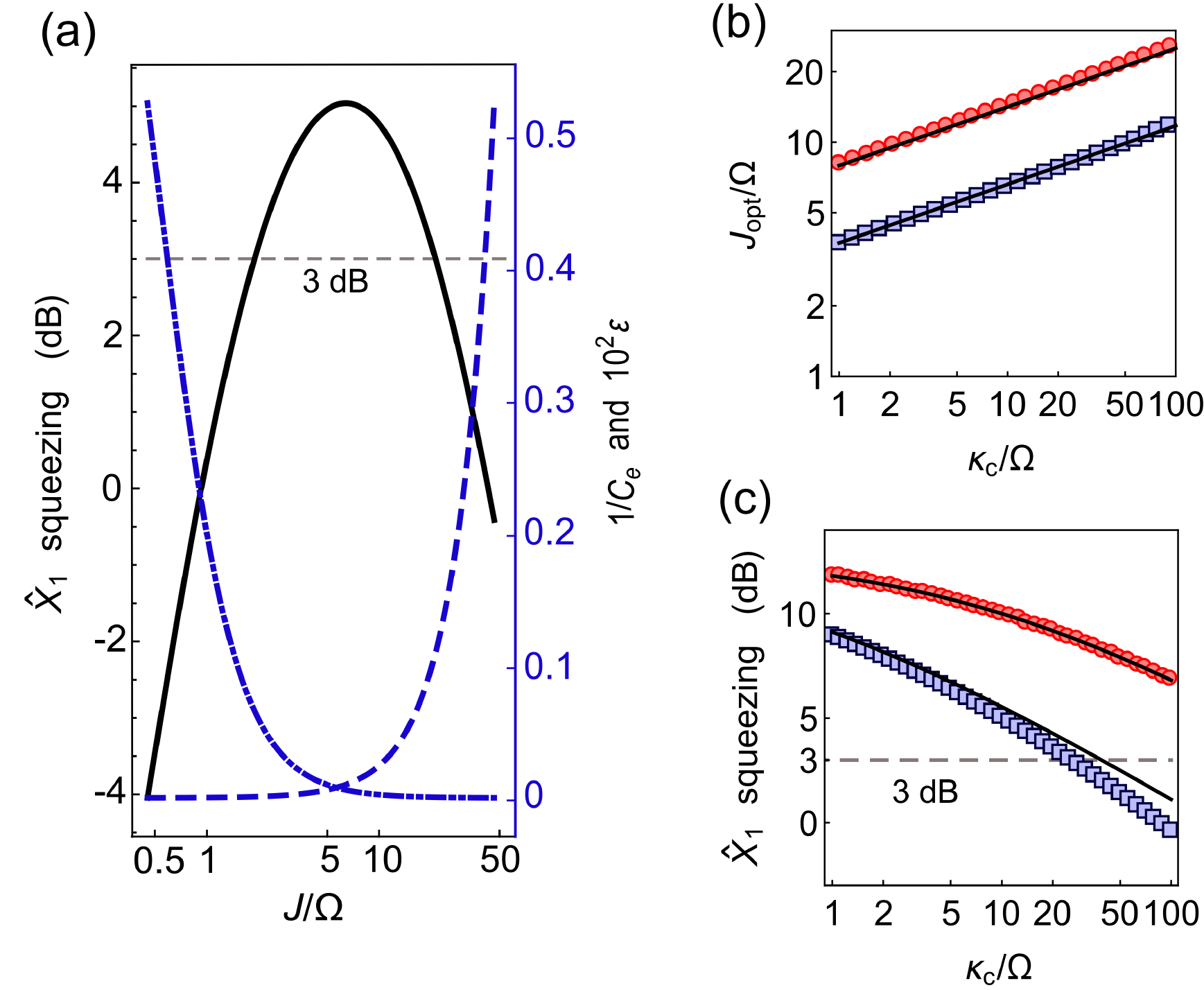}
\end{centering}
\caption{(Color online) (a) Mechanical squeezing vs. $J$ for optimized $r$.
The black solid line is the result of Eq.~(\ref{ORR}). The dashed and dash-dotted lines show the corresponding values of $1/C_{\rm e}$ and $\varepsilon$, see Eqs.~(\ref{Ce}) and (\ref{eps_pm}). Here $\kappa _{c}/\Omega =10$ and $n_{\text{th}}=10$. (b)
Optimal $J_{\text{opt}} $, with black lines from the approximate analytical result Eq.~(\ref{Jopt}). (c) Optimized squeezing, with black lines from the approximate analytical result Eq.~(\ref{optvar}). In both panels (b) and (c), the red circles ($n_{\text{th}}=0$) and blue squares ($n_{\text{th}}=10$) are obtained by minimizing Eq.~(\ref{ORR}) with respect to $J$. Other parameters: $\kappa
/\Omega =1/10$, $\gamma /\Omega =10^{-5}$, $G_{-}/\Omega =1/10$.}
\label{bound}
\end{figure}

The physics of optimization over the ratio $r$, discussed in the previous Sec.~\ref{sec:optimal_r}, is similar to the resolved sideband regime~\cite{Kronwaldpra2013,YDWang2013prl}. In the unresolved-sideband case, the coupling strength $J$  between the main and auxiliary cavities represents an additional crucial parameter for the design of the engineered reservoir. Evidently from Fig.~\ref{fig_schematics} and our previous discussion, a non-zero $J$ is able to mitigate the effect of unwanted counter-rotating terms. In particular, when $J$ is very large the spectrum reflects three well-separated hybridized modes, of which the one at $\omega =0$ is very sharp, i.e., leads to a small values of $\varepsilon$ (see Fig.~\ref{spectrum}). However, in the regime of large $J$ this central peak is mainly due to a superposition of auxiliary cavity modes, thus is very weakly coupled to the mechanical element and becomes ineffective in squeezing its thermal state. As a consequence, the variance has a non-monotonic dependence on $J$ and attains the smallest value at an optimal coupling $J_{\rm opt}$, see Fig.~\ref{bound}(a) for a concrete example.

Mathematically, the existence of such optimal point is indicated by Eqs.~(\ref{S0}) and (\ref{epsilon_largeJ}). As we have discussed in detail, Eq.~(\ref{epsilon_largeJ}) describes the decrease of $\varepsilon$ by increasing $J$,  which at moderate $J$ is beneficial to overcome the condition of non-resolved sidebands and obtain a larger mechanical squeezing. The strong decrease of $\varepsilon$ with $J$ is shown by the dot-dashed curve of Fig.~\ref{bound}(a). Eventually, $\varepsilon$ saturates to a small finite value when $J \gg \sqrt{\kappa \kappa_c}$. On the other hand, the dashed curve of Fig.~\ref{bound}(a) shows a strong decrease of the effective cooperativity $C_{\rm e}$ at large $J$, which can be understood from Eq.~(\ref{S0}): A large $J$ suppresses the spectral density at $\omega=0$ ($S_\mathrm{op}(0) \propto J^{-2}$ for $J \gg \Omega \sqrt{\kappa_c/\kappa} $, supposing $\Omega \gg \kappa$) and the decrease of $S_\mathrm{op}(0)$ implies a vanishing effective cooperativity, since $C_{\rm e} \propto S_\mathrm{op}(0)$. Therefore, increasing $J$ will eventually reduce the degree of squeezing, despite the tiny $\varepsilon$.

For a more quantitative analysis we resort to Eq.~(\ref{avl}), where the strengths of the laser drives are optimized, and consider the limit of $\Omega \gg \kappa$. We obtain the following approximation for $C_{\rm e}$:
\begin{equation}
C_{\rm e} \simeq C \left( 1+\frac{J^2 \kappa}{2\Omega^2 \kappa_c}\right)^{-1},
\end{equation}
where $C = 4 G_-^2/( \gamma\kappa_c)$ is the standard cooperativity. Furthermore, $\varepsilon$ is well described by Eq.~(\ref{epsilon_largeJ}). Performing these approximations in the first term of Eq.~(\ref{avl}) yields:
\begin{equation}\label{dX2_simplified}
\left\langle \Delta\hat{X}_{1}^{2}\right\rangle_{r_\text{opt}}\simeq
\frac12 \sqrt{C_{\rm th}^{-1}\left( 1+\frac{J^2 \kappa}{2\Omega^2 \kappa_c}\right)+\frac{\kappa \kappa_c}{2J^2}+\frac{\kappa^2}{4\Omega^2}},
\end{equation}
where the `thermal' cooperativity is defined as
\begin{equation}
C_{\rm th} = \frac{C}{2n_{\rm th}+1}.
\end{equation}
Note that the second contribution of Eq.~(\ref{avl}) was omitted: since we are interested in reaching a small variance, we need $n_{\text{th}}/(2C_{\text{e}}) \ll 1$. Then, the first term of Eq.~(\ref{avl}) is larger than $\sqrt{n_{\text{th}}/(2C_{\text{e}})} $, thus becomes the dominant one in this regime. Note also that Eq.~(\ref{dX2_simplified}) recovers Eq.~(\ref{bound1}) in the limit of infinite cooperativity.

Performing the optimization of Eq.~(\ref{dX2_simplified}) with respect to $J$ we finally obtain the maximum achievable squeezing:
\begin{equation}
\langle \Delta\hat{X}_{1}^{2}\rangle_{\rm opt} \simeq \frac12 \left(\sqrt{ C_{\rm th}^{-1}}+ \frac{\kappa}{2\Omega}\right ), \label{optvar}
\end{equation}
with both $r$ and $J$ optimized, and the optimal coupling
\begin{equation}
J_{\rm opt} \simeq   \left(C_{\rm th}\right)^{1/4}\sqrt{\Omega \kappa_c}. \label{Jopt}
\end{equation}
As shown in Fig.~\ref{bound}(b) and (c), these approximations are able to descibe accurately the numerical results. Furthermore the compact result of Eq.~(\ref{optvar}) highlights the two limiting factors of the squeezing protocol: the first is the thermal cooperativity and the second is due to the finite line width of the auxiliary cavities. These two sources of imperfection contribute to the minimum achievable variance in an additive way.

We also note that for typical system parameters the factor $\left(C_{\rm th}\right)^{1/4}\sqrt{\Omega /\kappa_c}$ is of order unity, so the optimal $J$ is of the same order of $\kappa_c$. In this regime, the corresponding optical spectrum is neither `Autler-Townes' nor `EIT'. Instead it shows three gentle peaks as the dot-dashed lines of Fig.~\ref{spectrum}(b).

\section{Discussion and Conclusions}\label{sec:imperfections}

\begin{figure}
\begin{centering}
\includegraphics[width=0.45\textwidth,angle=0]{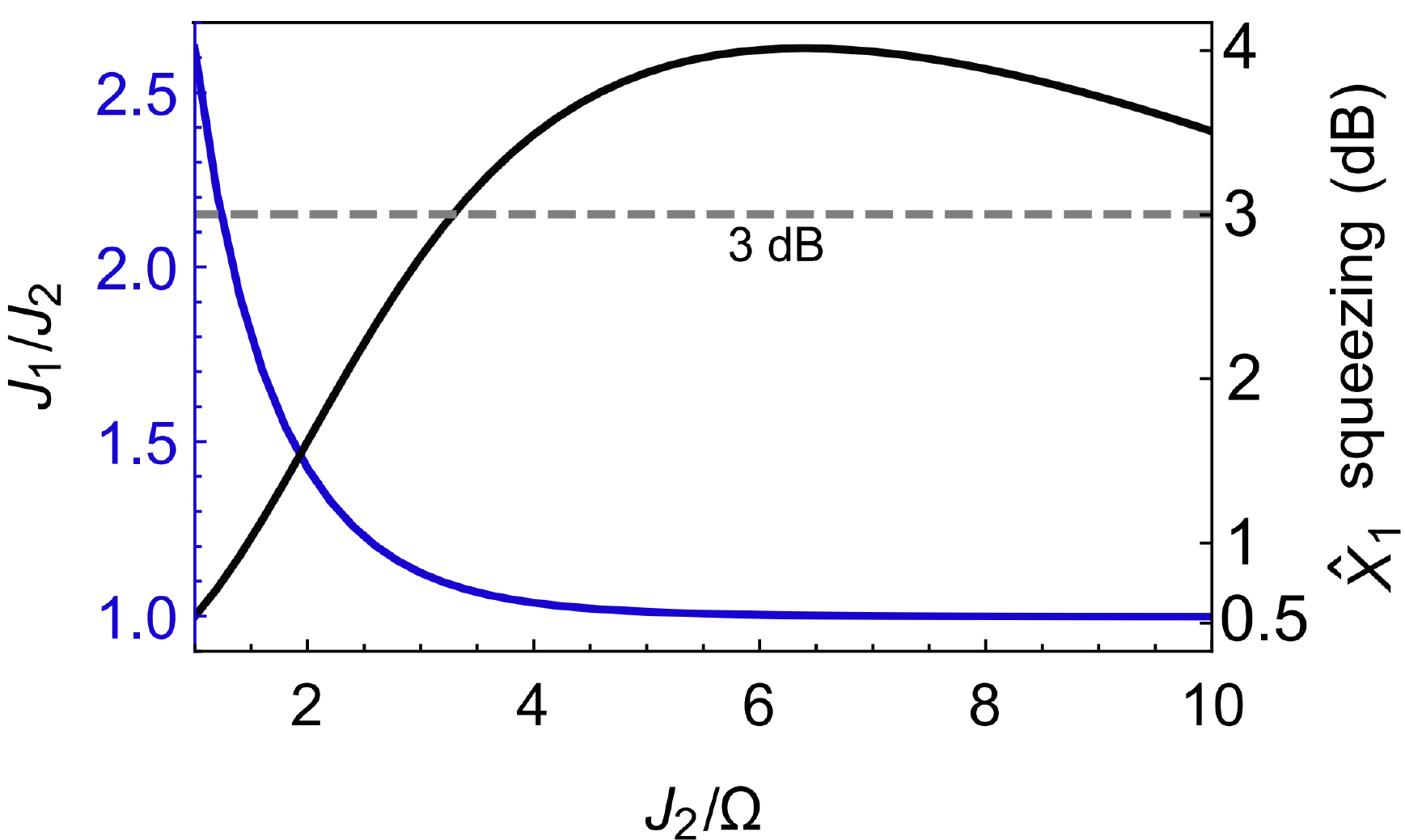}
\end{centering}
\caption{(Color online) The optimal ratio of $J_{1}/J_{2}$ (blue line) and the corresponding
mechanical squeezing (black line) vs. $J_2$. Other parameters are: $G_{-}/\Omega =1/10$,
$G_{+}=4/5G_{-}$, $\kappa _{c}/\Omega=10,\kappa /\Omega =1/5,\gamma /\Omega =1/5$
and $n_{\text{th}}=10$.
}%
\label{fig:optJ}%
\end{figure}

In previous sections, the two couplings between the main and auxiliary cavities have been assumed to
be equal, such that the optical spectrum $S_{\text{op}}\left(\omega\right)$ peaks at the cavity frequency (or $\omega=0$ in the
rotating frame). This is generally the optimal setting except for small $J_{1}/\Omega$ and $J_{2}/\Omega$, where the effect of
counter-rotating terms are comparable to the resonant terms. As shown Fig.~\ref{fig:optJ}, $J_{1}>J_2$ can suppress heating and enhance cooling to benefit squeezing. However, to achieve large squeezing, large values of $J_{1}$ and $J_{2}$
are desirable (see the black curve of Fig.~\ref{fig:optJ}) and, in this regime, the optimal choice of $J_{1}/J_2$ remains at the symmetric setting assumed in previous discussions.

Regarding the realization of the proposed setup, parameters we used in this paper about the optomechanical system and the high-finesse optical cavities are feasible with current technology, especially in photonic crystal nanobeams~\cite{JChan2011nat,OPainter2017natphys}. The only element which is not common is a strong coupling between the optomechanical system and the auxiliary cavities.
However, strong coupling between two optical cavities as large as $25$~GHz has already been realized in photonic crystal nanobeams~\cite{Susumu2012natpho}. In microtoroid system, optical cavities can also be coupled and the coupling strength can be sufficiently large~\cite{minxiaoprj2017} to reach $J>\Omega$.

In summary, we have shown that, for an optomechanical system in the unresolved
sideband regime driven with a two-tone laser, mechanical squeezing can still be
achieved with an improved version of reservoir engineering: the main cavity is coupled to two
auxiliary ones with carefully designed parameters. The role of
these additional cavities is to modulate the optical spectrum and suppress the
unwanted counter-rotating processes. The underlying mechanism is a quantum
interference effect analogous to EIT in atomic physics, and can lead to strong mechanical
squeezing (beyond 3 dB).

\begin{acknowledgments}
YDW acknowledges support from NSFC (Grants No.~11574330 and No.~11434011), MOST (Grant No.~2017FA0304500) and the Strategic Priority Research Program of Chinese Academy of Sciences (Grant No.~XDB23000000). SC acknowledges support from NSFC (Grants No.~11574025, No.~U1530401, No.~11750110428, and No.~1171101295). We also thank R. Fazio and G. C. La Rocca for helpful discussions.
\end{acknowledgments}

\appendix

\section{Linearization of the Hamiltonian}

\label{app_linearization}

We discuss here the derivation of the linearized Hamiltonian Eq.~(\ref{formula_ILH}). By introducing the displacement transformations mentioned in the main text: $\hat{a}=\alpha+\hat{d}$ , $\hat{a}_{1,2}=\alpha_{1,2}+\hat{d}_{1,2}$, and $\hat{b}=\beta+\hat{d}_{\rm m}$, the Langevin equations from the original Hamiltonian Eq.~(\ref{OH}) are as follows:
\begin{align}\label{Langevin_eq}
&\dot{\hat{d}}_{\rm m}   =-i\Omega\hat{d}_{\rm m}-\frac{\gamma}{2}\hat{d}_{\rm m}-ig(
\alpha^{\ast}\hat{d}+\alpha\hat{d}^{\dagger})
-ig\hat{d}^{\dagger}\hat{d}-\sqrt{\gamma}\hat{d}_{{\rm m},\text{in}}, \nonumber \\
&\dot{\hat{d}}   =-i\left(\omega_{c}+g(  \beta+\beta^*)\right)\hat{d}-\frac{\kappa_{c}}{2}\hat
{d}-ig\alpha(  \hat{d}_{\rm m}+\hat{d}_{\rm m}^{\dagger}) \nonumber\\
& \qquad ~~-ig\hat{d}(  \hat{d}_{\rm m}^{\dagger}+\hat{d}_{\rm m})
-iJ_{1}\hat{d}_{1}-iJ_{2}\hat{d}_{2}-\sqrt{\kappa_{c}}\hat
{d}_{c,\text{in}}, \nonumber  \\
&\dot{\hat{d}}_{i}    =-i\omega_{i}\hat{d}_{i}-\frac{\kappa_{i}}{2}%
\hat{d}_{i}-iJ_{i}\hat{d}-\sqrt{\kappa_{i}}\hat{d}_{i,\text{in}},
\end{align}
where the coherent amplitudes satisfy:
\begin{align}\label{amplitudes}
\dot{\beta} &= -i \Omega \beta -\frac{\gamma}{2}\beta -ig |\alpha|^2, \nonumber \\
\dot{\alpha}  &  =-i\omega_{c}\alpha-\frac{\kappa_{c}}{2}\alpha
-iJ_{1}\alpha_{1}-iJ_{2}\alpha_{2}-ig \alpha (\beta+\beta^*)\nonumber \\
&  \text{ \ \ }-i\left(  \alpha_{+}e^{-i\omega_{+}t}+\alpha_{-}e^{-i\omega
_{-}t}\right)  , \nonumber \\
\dot{\alpha}_{i}  &  =-i\omega_{i}\alpha_{i}-\frac{\kappa_{i}}{2}\alpha
_{i}-iJ_{i}\alpha,
\end{align}
In Eq.~(\ref{Langevin_eq}), $\hat{d}_{i,\text{in}}$ are white noise operators with correlation functions $\langle \hat{d}_{i,\text{in}}(  t)  \hat{d}_{i,\text{in}}^{\dagger}(t')  \rangle =\delta(  t-t')$ for the cavity modes ($i=c,1,2$), and $\langle \hat{d}_{{\rm m},\text{in}}^{\dagger}(  t')  \hat{d}_{{\rm m},\text{in}}(t)
\rangle =n_{\text{th}}\delta(t-t')  $, $\langle \hat {d}_{{\rm m},\text{in}}(  t)  \hat{d}_{{\rm m},\text{in}}^{\dagger}( t')  \rangle =(  n_{\text{th}}+1)\delta(  t-t')  $ for the mechanical bath, where $n_{\text{th}}$ is the thermal phonon number. All other noise correlation functions are zero.

By neglecting in Eq.~(\ref{Langevin_eq}) the small nonlinear terms and frequency shift of the main cavity $\delta \omega_c = g (\beta+\beta^*)$, the approximate Langevin equations define the following linearized Hamiltonian:
\begin{align} \label{HL_orig_frame}
\hat{H}_{\text{L}}  &  =\omega_{c}\hat{d}^{\dag}\hat{d}+\Omega\hat{d}_{\rm m}%
^{\dagger}\hat{d}_{\rm m}+g(\alpha^{\ast}\hat{d}+\alpha
\hat{d}^{\dagger})  (  \hat{d}_{\rm m}^{\dagger}+\hat{d}_{\rm m}) \nonumber\\
&  \text{ \ \ }+\sum_{i=1,2}\left(  \omega_{i}\hat{d}_{i}^{\dagger}\hat{d}%
_{i}+J_{i}(\hat{d}^{\dagger}\hat{d}_{i}+\hat{d}\hat{d}_{i}^{\dagger})\right)
+\hat{H}_{\rm env},
\end{align}
which is still written in the original frame. To obtain Eq.~(\ref{formula_ILH}) we should consider the explicit time-dependence of $\alpha$. To lowest order in $g$, Eq.~(\ref{amplitudes}) gives:
\begin{equation}\label{alpha_sol0}
\alpha \simeq \bar{\alpha}_{+} e^{-i\omega_{+}t}+\bar{\alpha}_{-}e^{-i\omega_{-}t},
\end{equation}
where:
\begin{equation}\label{baralpha}
\bar{\alpha}_{\pm} =\frac{\alpha_\pm}{\omega_\pm -\omega_c + i\frac{\kappa_c}{2}-\sum_{i=1,2}\frac{J_i^2}{ \omega_{\pm}-\omega_{i}+i\frac{\kappa_{i}}{2}}}.
\end{equation}
By defining the the many-photon couplings $G_\pm = g \bar{\alpha}_\pm$, which for definiteness we assume real (by a proper choice of the drive phases), and transforming Eq.~(\ref{HL_orig_frame}) to  an interaction picture with respect to $\omega_{c}(\hat{d}^{\dagger}\hat{d}+\hat{d}_1^{\dagger}\hat{d}_1+\hat{d}_2^{\dagger}\hat{d}_2)+\Omega \hat{d}_{\rm m}^{\dagger}\hat{d}_{\rm m}$, we finally obtain Eq.~(\ref{formula_ILH}) of the main text.

For completeness, we also give below the leading-order solutions for the classical amplitudes of the auxiliary cavities and mechanical mode:
\begin{align}
&\alpha_{i} \simeq J_{i}\left(\frac{\bar{\alpha}_{+}e^{-i\omega_{+}t}}{ \omega
_{+}-\omega_{i}+i\frac{\kappa_{i}}{2} }+\frac{\bar{\alpha}%
_{-}e^{-i\omega_{-}t}}{ \omega_{-}-\omega_{i}+i\frac{\kappa_{i}}%
{2} } \right), \label{alphai_sol0} \\
&\beta \simeq -g\frac{\bar{\alpha}_+^2+\bar{\alpha}_-^2}{\Omega -i \frac{\gamma}{2}}
+g\bar{\alpha}_+\bar{\alpha}_- \left(\frac{e^{-2i\Omega t}}{\Omega + i \frac{\gamma}{2}}
-\frac{e^{2i\Omega t}}{3\Omega -i \frac{\gamma}{2}} \right), \label{beta_sol1}
\end{align}
where the latter result is obtained by inserting Eq.~(\ref{alpha_sol0}) in the equation for $\dot\beta$ and using $\omega_\pm = \omega_c \pm \Omega$. The time-dependent contribution is due to the oscillation of the cavity intensity induced by the beat note between the two drives.

It is also worth mentioning that the above approximations require a sufficiently small drive strength, as can be seen by considering the corrections to the leading-order solution. Approximating the nonlinear term $-ig \alpha(\beta+\beta^*)$ in Eq.~(\ref{amplitudes}) through Eqs.~(\ref{alpha_sol0}) and (\ref{beta_sol1}), it is easily seen that additional Fourier components at $\omega_c \pm 3\Omega$ appear in the solution of $\alpha$, besides corrections at the original drive frequencies $\omega_c \pm \Omega$. To estimate the size of these corrections, we rely on Eq.~(\ref{baralpha}) and  $\kappa_i < \Omega < \kappa_c$ (the first inequality is necessary to achieve squeezing beyond 3 dB, see Sec.~\ref{sec:imperfections}) to estimate $\bar{\alpha}_\pm \sim \alpha_\pm / \max[\kappa_c,J_{1,2}^2/\Omega] $. Together with $\beta \sim g \bar\alpha_\pm^2/\Omega$, this gives:
\begin{equation}
g \alpha(\beta+\beta^*)\sim  \left\{ \frac{G_\pm^2}{\max[\Omega \kappa_c,J_{1,2}^2]}\right\} \alpha_\pm .
\end{equation}
Since $\alpha_\pm $ is the amplitude of the original drive, the factor in the curly brackets should much smaller than one for our treatment to be valid:
\begin{equation}\label{small_factor}
G_\pm \ll \max[\sqrt{\Omega \kappa_c},J_{1,2}].
\end{equation}

In practice, the condition Eq.~(\ref{small_factor}) is not very restrictive. In the main text, we generally assume $G_- \ll \Omega $ in giving explicit numerical results  (note that $\Omega<\kappa_c$, due to the unresolved-sideband regime). Furthermore, the optimal point of Eq.~(\ref{Jopt}) is in a regime of large $J$, with $J^2 =\sqrt{C_{\rm th}} \Omega \kappa_c \gg \Omega \kappa_c$. In this case, Eq.~(\ref{small_factor}) is much less restrictive than $G_- \ll \sqrt{\Omega \kappa_c}$.

\section{Maximally squeezed quadrature}

\label{appendix_maxsqueezing}

We consider the variance of $\hat{X}_{\theta}=\hat{X}_{1}\cos\theta+\hat{X}_{2}\sin\theta$, where $\hat{X}_{1}=(  \hat{d}_{\rm m}+\hat{d}_{\rm m}^{\dag})  /\sqrt{2}$ and $\hat{X}_{2}=i(  \hat{d}_{\rm m}^{\dag}-\hat{d}_{\rm m})  /\sqrt{2}$,
over a general squeezed vacuum state, given by $\left\vert r,\beta\right\rangle =\exp
[r(e^{i\beta}\hat{d}_{\rm m}^{2}-e^{-i\beta}\hat{d}_{\rm m}^{\dag2})/2]\left\vert
0\right\rangle $:
\begin{align}\label{Xtheta}
\langle\Delta\hat{X}_{\theta}^{2}\rangle &  =\left(  \sinh^{2}r-\cos\beta
\cosh^{2}r\sinh^{2}r+\frac{1}{2}\right)  \cos^{2}\theta\nonumber\\
&  \text{ \ \ }+\left(  \sinh^{2}r+\cos\beta\cosh^{2}r\sinh^{2}r+\frac{1}%
{2}\right)  \sin^{2}\theta\nonumber\\
\text{ \ \ }  &  \text{\ \ \ }-\sin2\theta\sin\beta\cosh r\sinh r.
\end{align}
In our case, since the Bogoliubov mode Eq.~(\ref{Bmode}) has real coefficients, we should set $\beta=0$. Then, Eq.~(\ref{Xtheta}) simplifies to:
\begin{equation}
\langle\Delta\hat{X}_{\theta}^{2}\rangle=\frac12 \left( e^{-2r}\cos^{2}\theta+ e^{2r}\sin^{2}\theta\right),
\end{equation}
showing that the maximally squeezed quadrature is obviously $\hat{X}_{1}$ (i.e., $\theta=0$).

\section{Effective master equation}

\label{appendix_ME}
We start by transforming the linearized Hamiltonian Eq.~(\ref{formula_ILH}) to an interaction picture with respect to the optical modes:
\begin{align}
\hat{H}_{I}=  &  \hat{d}^{\dag}(t)(G_{+}\hat{d}_{\mathrm{m}%
}^{\dagger}+G_{-}\hat{d}_{\mathrm{m}})+\text{H.c.}\nonumber\\
&  +\hat{d}^{\dag}(t)(G_{+}\hat{d}_{\mathrm{m}}e^{-2i\Omega t}+G_{-}\hat
{d}_{\mathrm{m}}^{\dagger}e^{2i\Omega t})+\text{H.c.} . \label{EqC1}
\end{align}
Here $\hat{d}(t)= e^{i\hat{H}_{\rm BO}t}\hat{d}e^{-i\hat{H}_{\rm BO}t}$, where $\hat{H}_\mathrm{BO}$ is defined in Eq.~(\ref{HBO}). We then apply the usual Born-Markov approximations to derive an effective master equation for the reduced mechanical density operator $\hat{\rho}(  t) =\rm{Tr}_{\rm B }[  \hat{\rho}_{\rm tot}(  t)  ] $:
\begin{equation}
\frac{d\hat{\rho}(  t)  }{dt}=-\int_{0}^{\infty}\text{Tr}_{\rm{B}}\left[
\hat{H}^{I}(t)  ,\left[  \hat{H}^{I}(  t-s)
,\hat{\rho}(  t)  \otimes\hat{\rho}_{B}\right]  \right]  ds . \label{BM_approx}
\end{equation}
Notice that in the above equations we only include the optical cavities as environment of the mechanical mode. For now we have omitted the thermal bath, which is uncorrelated with the structured optical bath. Its effect will be included at the end. Explicitly evaluating Eq.~(\ref{BM_approx})  through Eq.~(\ref{EqC1}) gives:
\begin{align}
\frac{d\hat{\rho}\left(  t\right)  }{dt}   = & \gamma_{-}[\hat{d}_{\rm m}\hat{\rho}\left(  t\right) , \hat
{d}_{\rm m}^{\dag}] +\gamma_{+}[\hat{d}_{\rm m}^{\dag}\hat{\rho}\left(  t\right) , \hat{d}_{\rm m}]\nonumber\\
&+\gamma_{S}\left([\hat{d}_{\rm m}\hat{\rho}\left(  t\right) , \hat {d}_{\rm m}] +[\hat{d}_{\rm m}^{\dag}\hat{\rho}\left(  t\right) , \hat{d}_{\rm m}^{\dag}] \right) + { \rm H.c.},  \label{Meq_derived}
\end{align}
with the coefficients:%
\begin{align}
\gamma_{\pm}    = &\int_{0}^{\infty} ds
\left[\left(G_{+}^{2}e^{i\Omega s}+G_{-}^{2}e^{-i\Omega s}\right)\langle
\hat{d}(s)\hat{d}^{\dag}(0) \rangle  \right. \nonumber \\
& \qquad~~\left. +\left(G_{+}^{2}e^{-i\Omega s}+G_{-}^{2}e^{i\Omega s}\right)\langle
\hat {d}^{\dag}(s)\hat{d}(0) \rangle \right]
e^{\mp i\Omega s} , \nonumber \ \\
\gamma_{S}  = & G_{+}G_{-}  \int_{0}^{\infty}ds
\left( \langle \hat{d}^{\dag}(s)\hat{d}(0)\rangle +\langle\hat{d}(s)\hat{d}^{\dag}(0)\rangle \right) ,
\end{align}
Notice that, in obtaining Eq.~(\ref{Meq_derived}), we have neglected terms with an explicitly time-dependence of the type
$\exp[i\Omega t]$. These rapidly oscillating terms have a small effect, since the typical time scale of the intrinsic evolution $\tau_{S}\sim1/\Omega$ is much shorter than the time $\tau_{R}\sim1/\gamma$ over which $\hat{\rho}\left(  t\right)  $ varies appreciably. Finally, by defining $\Gamma_{i}   =2{\rm Re}[\gamma_{i}]$ and $\Upsilon_{i}={\rm Im}[\gamma_{i}]$ (with $i=\pm, S$), the master equation becomes more compact:%
\begin{align}
\frac{d\hat{\rho}(t)}{dt}  &  =-i[\hat{H}_{\text{LS}},\hat{\rho}(t)]+\Gamma_{-}%
\mathcal{D}(\hat{d}_{\rm m})\hat{\rho}+\Gamma_{+}\mathcal{D}(\hat{d}_{\rm m}^{\dag})\hat{\rho
}\nonumber\\
&  \text{ \ \ }+\Gamma_{S}\left(  \mathcal{D}_{S}(\hat{d}_{\rm m})\hat{\rho
}+\mathcal{D}_{S}(\hat{d}_{\rm m}^{\dag})\hat{\rho}\right)  , \label{appMe}%
\end{align}
with%
\begin{equation}
\hat{H}_{\text{LS}}=\Upsilon_{-}\hat{d}_{\rm m}^{\dag}\hat{d}_{\rm m}+\Upsilon_{+}\hat{d}_{\rm m}%
\hat{d}_{\rm m}^{\dag}+\Upsilon_{S}\left(  \hat{d}_{\rm m}^{2}+\hat{d}_{\rm m}^{\dag2}\right)  ,
\end{equation}
and%
\begin{align}
\Gamma_{+}  &  =G_{+}^{2}S_{\text{op}}\left(  0\right)  +G_{-}^{2}%
S_{\text{op}}\left(  -2\Omega\right)  +\gamma n_{\text{th}},\\
\Gamma_{-}  &  =G_{-}^{2}S_{\text{op}}\left(  0\right)  +G_{+}^{2}%
S_{\text{op}}\left(  2\Omega\right)  +\gamma\left(  1+n_{\text{th}}\right)
,\\
\Gamma_{S}  &  =G_{+}G_{-}S_{\text{op}}\left(  0\right)  .
\end{align}
Here we have also included the mechanical thermal bath, by adding the appropriate heating and cooling rates to $\Gamma_\pm$\color{black}. Neglecting the small effect of the Lamb shift, we obtain Eq.~(\ref{formula_master}) of the main text.

\section{Lindblad form of the master equation}

\label{appendix_DME}

To write Eq.~(\ref{formula_master}) explicitly in Lindblad form, we introduce a Bogoliubov mode $\hat{B}'$:
\begin{equation}
\hat{d}_{\rm m}=u\hat{B}'+v\hat{B}'^{\dagger}%
\end{equation}
where  $u$ and $v$ are supposed to be real ($u^{2}-v^{2}=1$). Then, Eq.~(\ref{formula_master}) can be rewritten as follows:%

\begin{align}\label{ME_cmode}
&\frac{d\hat{\rho}\left(  t\right)  }{dt}   =
\left( v^{2} \Gamma_{-}+u^{2}\Gamma_{+}+2uv\Gamma_{S}\right)
\mathcal{D}(  \hat{B}'^{\dagger})  \hat{\rho}\nonumber\\
&  \text{ \ \ }+\left( u^{2} \Gamma_{-}+v^{2}\Gamma_{+}+2uv\Gamma_{S}\right)
\mathcal{D}(  \hat{B}')  \hat{\rho}\nonumber\\
&  \text{ \ \ }+\left(  uv(  \Gamma
_{-}+\Gamma_{+})  +(  u^{2}+v^{2})\Gamma_{S}  \right)
\left(  \mathcal{D}_S(  \hat{B}') \hat{\rho} +\mathcal{D}_S(  \hat
{B}'^{\dagger})\hat{\rho}  \right).
\end{align}
The last line is zero for the following choice of $u$ and $v$:
\begin{equation}
u  =\frac{\Gamma_{s}}{b}\sqrt{\frac{2b}{  a-b}}, \qquad
v  =-\sqrt{\frac{a-b}{2b}},
\end{equation}
where the definitions of $a$ and $b$ are given after Eq.~(\ref{mode_c}). These results are in agreement with the main text and the rates in the first and second line of Eq.~(\ref{ME_cmode}) are the $\Gamma^{B'}_{\pm}$, given after Eq.~(\ref{mode_c}).

\section{Dependence of squeezing on $\varepsilon_\pm $ and $C_\text{e}$}

\label{app_dependence}

\begin{figure}[ptb]
\begin{centering}
\includegraphics[width=0.5\textwidth,angle=0]{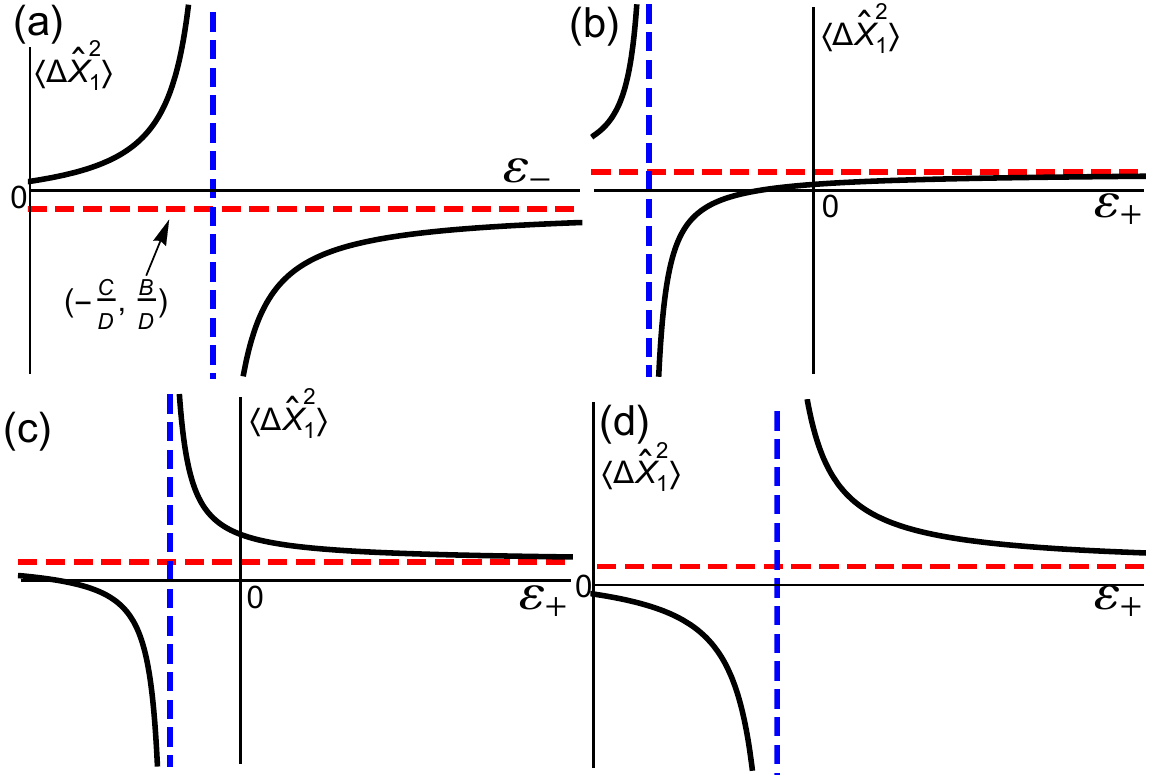}
\end{centering}
\caption{(Color online) Four general possibilities of mechanical variance
$\langle\Delta\hat{X}_{1}^{2}\rangle$ versus $\varepsilon_{\pm}$. See the main
text for a detailed explanation of the four cases. Although $\varepsilon_{\pm
}$ are positive quantities, to illustrate the functional dependence of
Eq.~(\ref{svariance1}) we extended panels (b) and (c) to $\varepsilon
_{+} <0$.}
\label{var_4cases}
\end{figure}

The parameters $\varepsilon_\pm $, representing the strength of the counter-rotating terms, play an important role in the generation of squeezing. Supposing that the other parameters $\zeta$, $C_{\mathrm{e}}$, and $n_{\mathrm{th}}$ are held constant, Eq.~(\ref{svariance1}) is of the simple form $(A+Bx)/(C+Dx)$ (where $x=\varepsilon_-$ or $\varepsilon_+$) and leads to the four cases illustrated in Fig.~\ref{var_4cases}. From cases (a) and (c) two necessary conditions for squeezing will be derived, see Eqs.~(\ref{max_nth}) and (\ref{condition2}).

Case (a) of Fig.~\ref{var_4cases} occurs by fixing $\varepsilon_+$ and considering $x=\varepsilon_-$ as a variable. The asymptotic value is $B/D = -1/2$ and it is easy to see from Eq.~(\ref{svariance1}) that the pole is at $-C/D >0$. Since the physically meaningful region is on the left side of the pole, it is indeed true that the variance is monotonically increasing with $\varepsilon_-$. Mechanical squeezing is not possible unless the variance is smaller than 1/2 at $\varepsilon_-=0$, which leads to the following constraint on the thermal occupation:
\begin{equation}\label{max_nth}
n_{\text{th}}<   \frac{C_{\text{e}}(1-e^{-2\zeta})}{ 2 \cosh^{2}\zeta }  ,
\end{equation}

The other three cases (b), (c), and (d) correspond to $x=\varepsilon_+$ as a variable while fixing $\varepsilon_-$. The asymptotic value is $B/D = 1/2$, and the position of the pole is given by:
\begin{equation}
\frac{C}{D}= \frac{1+\left(1/C_{\text{e}}-\varepsilon_{-}\right)  \cosh^{2}\zeta}{\sinh
^{2}\zeta}. \label{r1}%
\end{equation}
Panels (b) and (c) assume $C/D>0$, i.e., $\varepsilon_- <  1/\cosh^{2}\zeta+1/C_{\rm e}$. At $\varepsilon_+=0$ the variance is:
\begin{equation}
\frac{A}{C}=\frac12 ~ \frac{e^{-2\zeta}+\left(  \varepsilon_{-}+\left(  1+2n_{\text{th}}\right)
/C_{\text{e}}\right)  \cosh^{2}\zeta}{1+\left(
1/C_{\text{e}}-\varepsilon_{-}\right)  \cosh^{2}\zeta}, \label{r3}%
\end{equation}
which is positive since $C/D>0$. Then there are two cases: $0<A/C<1/2$ is plotted in panel (b), where the variance monotonically decreases with $\varepsilon_{+}$ and the largest squeezing is achieved at $\varepsilon_+ =0$; $A/C>1/2$ is plotted in panel (c), where decreasing $\varepsilon_+$ leads to a larger variance. This dependence is opposite to what one would expect, however here $\langle \Delta \hat{X}_1^2 \rangle$ is always larger than $1/2$. Thus, the latter regime is not interesting for squeezing zero-point motion.

Following this discussion, we get another necessary condition for squeezing:
\begin{equation}\label{condition2}
\varepsilon_- < \frac{1-e^{-2\zeta}}{2\cosh^{2}\zeta}- \frac{n_{\rm th}}{C_{\text{e}}},
\end{equation}
which can be simply obtained from Eq.~(\ref{r3}) by setting $A/C<1/2$. Furthermore, the right hand side of Eq.~(\ref{condition2}) should be larger than zero (since $\varepsilon_-$ is always positive), which allows us to recover the bound on $n_{\rm th}$ given in Eq.~(\ref{max_nth}).

The last case to consider is $C/D<0$, which leads to panel (d). Comparing Eqs.~(\ref{r1}) and (\ref{r3}), one can see that $A/C$ is negative. Therefore, Eq.~(\ref{svariance1}) implies either an unphysical negative value (on the left of the pole) or no squeezing at all (on the right side).

In summary we find that, in all cases where squeezing is possible, the variance is reduced by decreasing $\varepsilon_\pm$. Following a similar proof we can show that, when the other parameters are fixed, increasing $C_\text{e}$ always reduces the variance.

\bibliographystyle{apsrev4-1}
\bibliography{squeezing}

\bigskip

\end{document}